\documentstyle[preprint,aps,prl]{revtex}

\begin{document}
\draft

\title{PEAK VALUES OF CONDUCTIVITY IN INTEGER AND FRACTIONAL
QUANTUM HALL EFFECT}

\author{L. P. Rokhinson, B. Su and V. J. Goldman}

\address{Department of Physics, State University
of New York, Stony Brook, NY 11794-3800}

\date{Solid State Communications, {\bf 96} 309 (1995)%
\newline Received by E. E. Mendez June 5, 1995)}

\maketitle

\begin{abstract}

The diagonal conductivity $\sigma_{xx}$ was measured in the Corbino
geometry in both integer and fractional quantum Hall effect (QHE).
We find that peak values of $\sigma_{xx}$ are approximately equal for
transitions in a wide range of integer filling factors $3<\nu<16$, as
expected in scaling theories of QHE.  This fact allows us to compare
peak values in the integer and fractional regimes within the framework
of the law of corresponding states.

\vskip 1.5em


\end{abstract}

\newpage

Since the discovery of the quantum Hall effect (QHE) considerable
attention has been devoted to the transition region between quantum Hall
plateaus. Experiments\cite{narita82} on low mobility samples
showed that peak values of the diagonal conductivity are Landau-level
dependent ($\sigma^{peak}_{xx}\propto\nu$), as predicted
by Ando and Uemura\cite{ando74}.  Similar dependence was
observed\cite{coleridge94} in high mobility samples for transitions at
high filling factors.  However, the scaling theory of
QHE\cite{levine83,khmelnitskii83} predicts that peak values of
$\sigma_{xx}$ in the integer QHE (IQHE) should be universal,
independent of the filling factor.  A crossover from the Ando to the
scaling regime should occur at magnetic fields such that the magnetic
length is comparable to the range of the disorder potential.  In the
scaling regime the extended states exist only near a single energy in
each Landau level; a phase transition occurs when the chemical
potential crosses from the localized states of $i$-th Landau level
($\nu=i$ QHE plateau) through the extended states (peak of
$\sigma_{xx}$ at $\nu\approx i+1/2$) to the localized states of the
($i+1$)-st Landau level ($\nu=i+1$ QHE plateau).

The scaling regime
was extensively studied theoretically, however universality of peak
values of $\sigma_{xx}$ has not been observed
experimentally\cite{note7}.  Recently, several authors calculated peak
values for both IQHE \cite{note10,kivelson92,huo93,dykhne94} and
fractional QHE (FQHE)\cite{kivelson92,dykhne94,chklovskii93a} regimes.
In this paper we present an experimental study of the peak values of
$\sigma_{xx}$.  We have observed that the peak values of $\sigma_{xx}$
are nearly equal for a wide range of transitions between IQHE states.
This observation allows us to compare peak values of $\sigma_{xx}$ in
the integer and fractional QHE regimes for the same sample. We show
that transitions between FQHE states can be successfully mapped onto
transitions between IQHE states, according to the law of corresponding
states\cite{jain90}.

We chose a Corbino geometry (see inset in Fig. 1) in order to
avoid edge channels connecting sample contacts.  In samples with
Hall-bar geometry the longitudinal resistance $R_{\text{L}}$ is not
proportional to the local resistivity $\rho_{xx}$ because of non-local
transport\cite{mceuen90}.  This effect makes measurements of
$R_{\text{L}}$ geometry and sample dependent\cite{note8}.  In
the Corbino geometry the measured two-terminal resistance
$R_{\text{2T}}$ is inversely proportional to the local conductivity:
$R_{2T}=\Box/\sigma_{xx}$, where $\Box=1/2\pi\ln(r_2/r_1)$ is a
geometrical factor (``number of squares'').  We present all our data
in terms of $\sigma_{xx}$, calculated according to this formula from
measured $R_{\text{2T}}$.
Samples were fabricated from high mobility
($1.5<\mu<2.0\times10^6\text{ cm}^2\text{/V s}$) GaAs/AlGaAs
heterojunction wafers. Corbino geometry was defined by circular Ohmic
contacts with the inner radius $0.2\leq r_1\leq0.6$ mm and the
outer radius $r_2=1.5$ mm.  Two-dimensional electron systems (2DES)
were prepared by illuminating samples with a red LED.  Temperature was
measured with a calibrated Ruthenium Oxide chip resistor and the
absolute values are believed to be accurate to 5\%. Measurements were
done using standard lock-in technique at 2.5 Hz with an applied current
50 pA rms; no heating effects were observed at this current.

Representative magnetoconductivity data are shown in Figs. 1
and 2.  In IQHE (Fig. 1)
minima in $\sigma_{xx}$ are well developed up to filling factor
$\nu=40$.  At a temperature $T<100$ mK we observe a remarkable result:
peaks of $\sigma_{xx}$ form an approximately flat region between
$\nu=3$ and 16; the values of $\sigma^{peak}_{xx}$ fall within the
dashed lines in Fig. 1, which are displaced by $\pm7\%$ from the
average value of 0.22 $e^2/h$.  The average peak value varies in
different samples between 0.20 and 0.35 $e^2/h$ at the lowest $T=13$
mK.  The range of $\nu$ where $\sigma^{peak}_{xx}$ is $\nu$-independent
also changes a little from sample to sample. The flat region is well
defined up to $T\sim40-100$ mK and disappears at higher temperatures
(compare data for 20 mK and 1.8 K in Fig. 1). At low fields ($\nu>20$),
values of $\sigma^{peak}_{xx}$ follow the Ando dependence
$\sigma^{peak}_{xx}\approx \beta {{e^2}\over{h}}\nu$ in the full
experimental temperature range (13 mK $\leq T\leq$ 1.6 K) with the
coefficient $\beta\approx0.1$.  The same value of $\beta$ is obtained
from the amplitude of Shubnikov - de Haas oscillations, following
Coleridge {\it et al.}\cite{coleridge94}.

The relation between the peak values of conductivity in
IQHE and FQHE is understood naturally within the framework of the ``law
of corresponding states''\cite{jain90}.  This law predicts
equivalence between two QHE systems with filling factors $\nu^*$ and
$\nu$ related through:
\begin{description}
\item i) Landau level addition transformation
$\nu^*\leftrightarrow\nu+1$;
\item ii) particle-hole conjugation
$\nu^*\leftrightarrow 1-\nu$;
\item iii) flux attachment transformation
$(\nu^{-1})^*\leftrightarrow\nu^{-1}+2 m$,
\end{description}
where $m$ is an integer. Universality of $\sigma^{peak}_{xx}$ in IQHE is a
manifestation of the first transformation  and was discussed above,
while iii) relates the system of composite fermions\cite{jain89} (CF) in
FQHE and electrons in IQHE.  The peak conductivity for CF,
$\sigma^{peak}_{cf}$, can be related to the experimentally measurable
value of $\sigma^{peak}_{xx}$ through equations derived in
Ref.\ \onlinecite{chklovskii93a}:
\begin{eqnarray}
\label{sFQHE}
&&\rho_{xx}={{h}\over{e^2}}{{\sigma_{cf}^{peak}}\over
{(\sigma_{cf}^{peak})^2+(\nu_{cf})^2}},\nonumber\\
\\
&&\rho_{xy}=-{{h}\over{e^2}}\biggl[{{\nu_{cf}}\over
{(\sigma_{cf}^{peak})^2+(\nu_{cf})^2}}+2\biggr].\nonumber
\end{eqnarray}
Here $\nu_{cf}$ is the filling factor of CF at which the transition
occurs, $\sigma^{peak}_{cf}$ is in units of $e^2/h$ and the
conductivity tensor can be obtained by matrix inversion from the
resistivity tensor.  Note that for the $m=1$ main FQHE
sequence\cite{jain92} $\nu=\nu_{cf}/(2\nu_{cf}+1)$; for example, the
transition between $\nu_{cf}=1$ and 2 for CF (at $\nu_{cf}\approx3/2$)
corresponds to the transition between $\nu=1/3$ and 2/5 for electrons
(at $\nu\approx3/8$)\cite{goldman90}.

Peak values for CF conductivity are expected\cite{chklovskii93a} to
be the same as for electrons in IQHE. Thus, Eq. \ref{sFQHE} relates
$\sigma^{peak}_{xx}$ in IQHE and FQHE regimes. Such comparison is
meaningful only because the peak $\sigma_{xx}$ values are nearly the
same in IQHE in our experiments.  We introduce
$\sigma_{\text{IQHE}}^{peak}=\ <\sigma_{xx}^{peak}>$, where averaging
is done over IQHE transitions for which $\sigma_{xx}^{peak}$ differs
less than 15\% from the mean value. The correspondence between IQHE and
FQHE is demonstrated in Fig.  2: values for open and solid circles are
calculated from Eq.  \ref{sFQHE} using experimental
$\sigma^{peak}_{cf}=\sigma_{\text{IQHE}}^{peak}=0.34\ e^2/h$ and
theoretical\cite{note10,kivelson92,huo93,dykhne94}
$\sigma^{peak}_{cf}=0.5\ e^2/h$ values, respectively.  There is no
adjustable parameters in this calculation; open circles fit the data
better, as could be expected.

Peak conductivity for CF $\sigma^{peak}_{cf}$ can be obtained directly
from the experimental FQHE conductivity data via Eq. \ref{sFQHE},
without considering {\it a priori} any particular relation between
$\sigma^{peak}_{cf}$ and $\sigma^{peak}_{\text{IQHE}}$. In Fig. 3 thus
determined $\sigma^{peak}_{cf}$ is plotted for several transitions
between FQHE states together with $\sigma_{\text{IQHE}}^{peak}$ for
IQHE regime in a wide temperature range.  Again, this comparison
involves no adjustable parameters; we interpret the close agreement of
the absolute values of $\sigma^{peak}_{cf}$ and
$\sigma_{\text{IQHE}}^{peak}$ as well as their very similar temperature
dependence as an experimental confirmation of the law of
corresponding states.

Although the peak values of $\sigma_{xx}$ for electrons in IQHE and CF
in FQHE are nearly the same, they are not saturating to 0.5 $e^2/h$ at
$T\rightarrow0$ (Fig. 3), as
expected\cite{note10,kivelson92,huo93,dykhne94} for non-interacting
particles.  In contrast, the peak conductivity decreases at low
temperatures.  The similarity in the $T$-dependence of peak
conductivities $\sigma^{peak}_{cf}$ and $\sigma_{\text{IQHE}}^{peak}$
in the full experimental range of temperatures suggests that the
underlying physics is essentially the same in both regimes.  In low
mobility samples a similar reduction of conductivity was attributed to
interaction effects\cite{paalanen83}. It should be noted that in QHE
regime weak localization is not expected to alter $\sigma_{xx}^{peak}$
because high $B$ breaks time reversal symmetry.

To summarize our results, we observed, for the first time, that the peak
values of $\sigma_{xx}$ are nearly equal for a wide range of
transitions between IQHE states as was predicted in the scaling
theories of QHE.  This observation allows us to compare peak values of
$\sigma_{xx}$ in the integer and fractional QHE regimes for the same
sample. We show that transitions between FQHE states can be
successfully mapped onto transitions between IQHE states, according to
the law of corresponding states. In other words, peak values of
diagonal conductivity for electrons in IQHE are the same as for
composite fermions in FQHE.

We are grateful to M. Shayegan for MBE material and to S. Sondhi, I. M.
Ruzin, D. B. Chklovskii and D. E.  Khmel'nitskii for interesting
discussions. This work was supported in part by NSF under Grant
No. DMR-9318749.


\begin{references}
\bibitem{note7}
The reason for experimental difficulties is that either high disorder in low
  mobility samples or FQHE in high mobility samples limit the range of filling
  factors with scaling behavior in the IQHE regime; this range may be too
  narrow to be observed.
\end{references}

\begin{figure}
\label{sigbi}
\caption{Magnetoconductivity in the IQHE regime at 20
mK (solid line) and 1.8 K (dotted line) for Sample A
($n=1.05\times10^{11}\ \text{cm}^{-2}$).  Horizontal solid line gives
the average peak conductivity $\sigma^{peak}_{\text{IQHE}}=0.22\ e^2/h$
(for $\nu$ between 3 and 16) and the dashed lines are offset $\pm 7\%$
from the average value.  Sample layout in the Corbino geometry is shown
in the inset.}
\end{figure}

\begin{figure}
\label{sigbf}
\caption{Magnetoconductivity in the FQHE regime for Sample B
($n=1.17\times10^{11}\ \text{cm}^{-2}$) at 14 mK (solid line).  Also
shown are peak values of conductivity $\sigma_{xx}$ calculated from
$\sigma^{peak}_{cf}$ for composite fermions using Eq.\ \protect%
\ref{sFQHE} with (i) theoretically expected value of $0.5\ e^2/h$
($\bullet$) and (ii) the experimentally obtained value
$\sigma^{peak}_{\text{IQHE}}=0.35\ e^2/h$ ($\circ$).}
\end{figure}

\begin{figure}
\label{sA-t}
\caption{The temperature dependence of the peak values
$\sigma_{\text{IQHE}}^{peak}$ for transitions between IQHE states
(solid lines) and of $\sigma^{peak}_{cf}$ for composite fermions for
transitions between FQHE states $1/3\leftrightarrow2/5$ ($\circ$) and
$2/5\leftrightarrow3/7$ ($\bullet$).  $\sigma^{peak}_{cf}$ is obtained
from the experimental data using Eq.\ \protect\ref{sFQHE}.}
\end{figure}

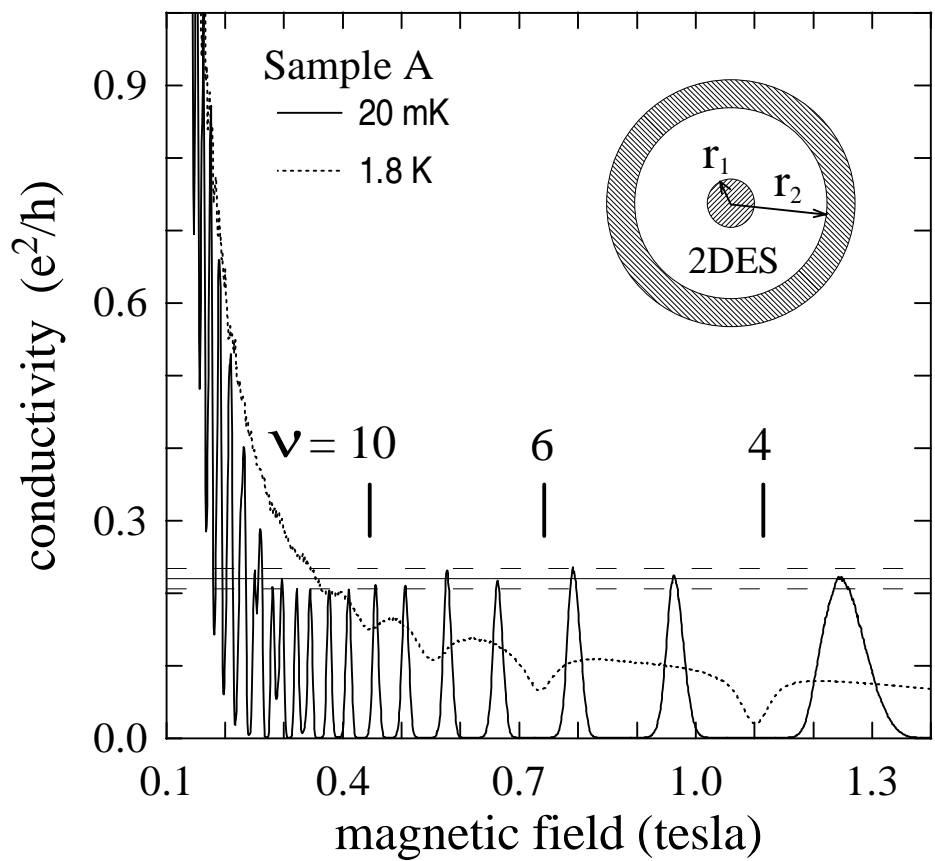
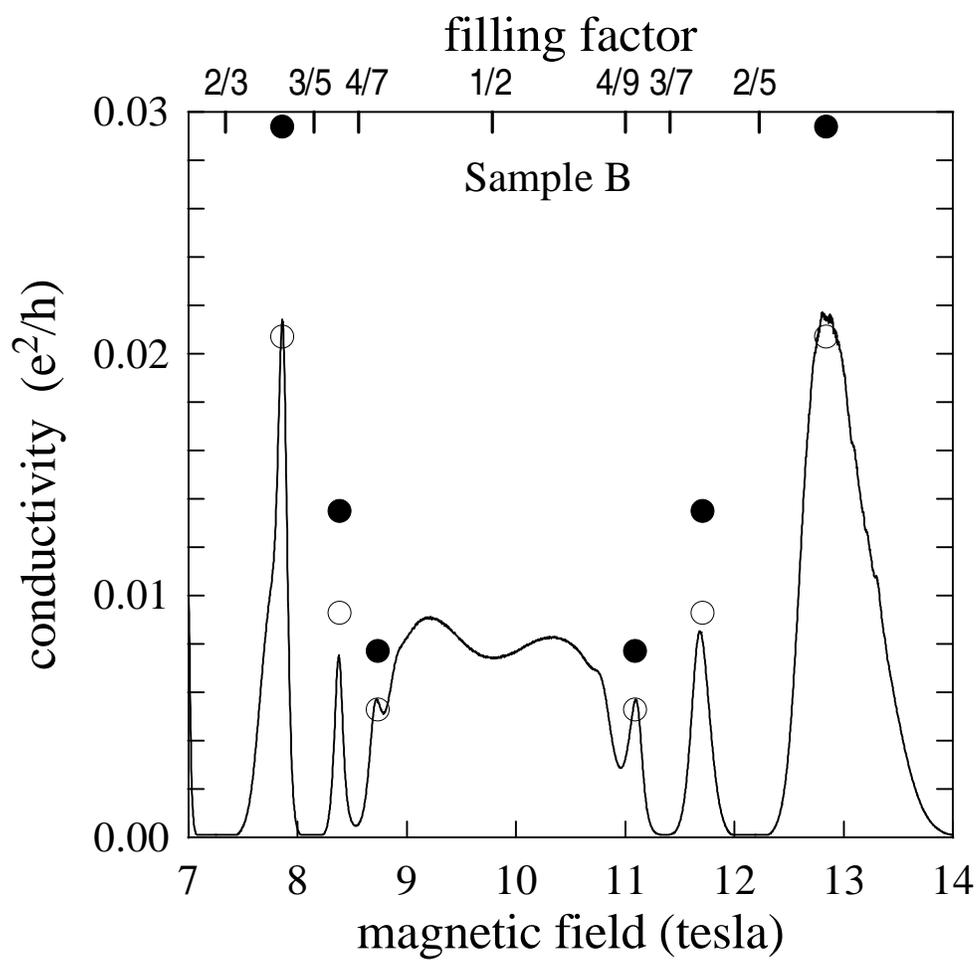
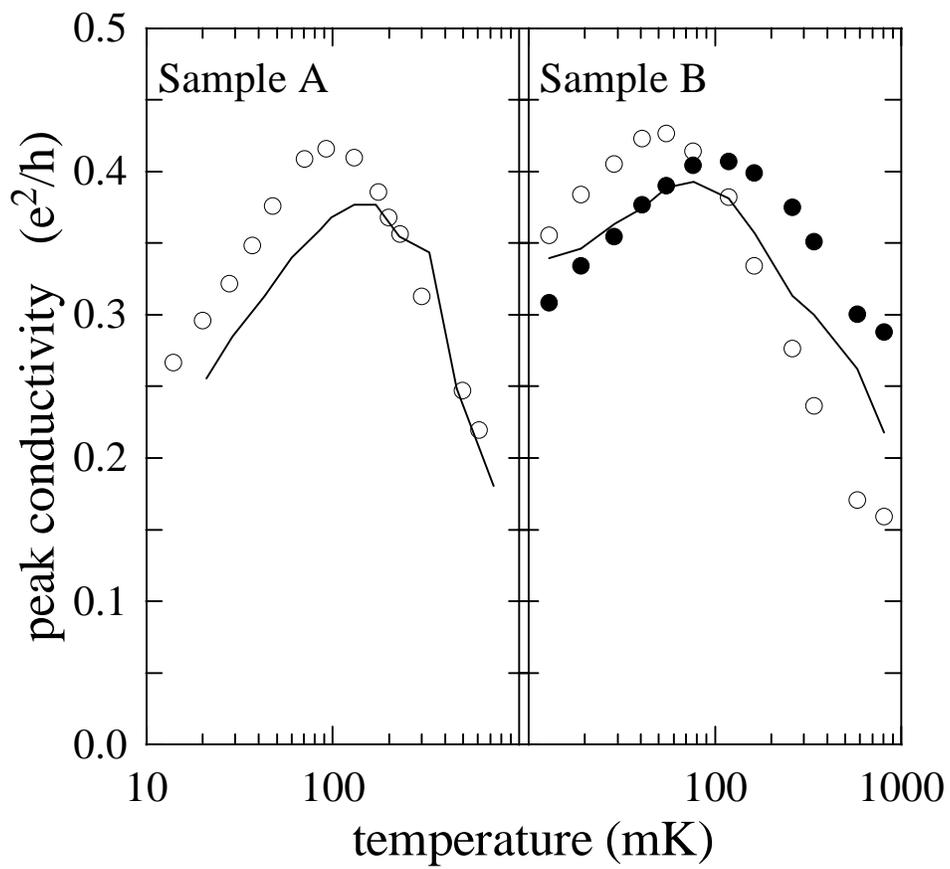
\end{document}